\documentclass[journal=jacsat,manuscript=article]{achemso}

\usepackage[version=3]{mhchem} % Formula subscripts using \ce{}
\usepackage{float}
\usepackage{comment}
\usepackage{amssymb}
\usepackage{xcolor}
\author{Ahmed Lafeef Ettapuram Naduvilepurayil}
\affiliation[Ariel University]
{Department of Electrical and Electronics Engineering, Ariel University, Ariel 40700, Israel}
\author{Harel Ginat}
\affiliation[Ariel University]
{Department of Electrical and Electronics Engineering, Ariel University, Ariel 40700, Israel}
\author{Fernando Lorén}
\affiliation[Universidad de Zaragoza]
{Instituto de Nanociencia y Materiales de Arag\'{o}n (INMA), CSIC-Universidad de Zaragoza, 50009 Zaragoza, Spain}
\alsoaffiliation[Universidad de Zaragoza]
{Departamento de F\'{i}sica de la Materia Condensada, Universidad de Zaragoza, 50009 Zaragoza, Spain}
\author{Luis Martín-Moreno}
\affiliation[Universidad de Zaragoza]
{Instituto de Nanociencia y Materiales de Arag\'{o}n (INMA), CSIC-Universidad de Zaragoza, 50009 Zaragoza, Spain}
\alsoaffiliation[Universidad de Zaragoza]
{Departamento de F\'{i}sica de la Materia Condensada, Universidad de Zaragoza, 50009 Zaragoza, Spain}
\author{Shmuel Sternklar}
\affiliation[Ariel University]
{Department of Electrical and Electronics Engineering, Ariel University, Ariel 40700, Israel}
\author{Yuri Gorodetski}
\email{yurig@ariel.ac.il}
\affiliation[Ariel University]
{Department of Electrical and Electronics Engineering, Ariel University, Ariel 40700, Israel}
\alsoaffiliation[Ariel University]
{Department of Mechatronics and Mechanical Engineering, Ariel University, Ariel 40700, Israel}
\title[Sagnac Paper]{Direct interferometric measurement of non-reciprocity induced by a plasmonic metasurface with false chirality}


\begin{document}

\begin{tocentry}
    \includegraphics[width=1\textwidth]{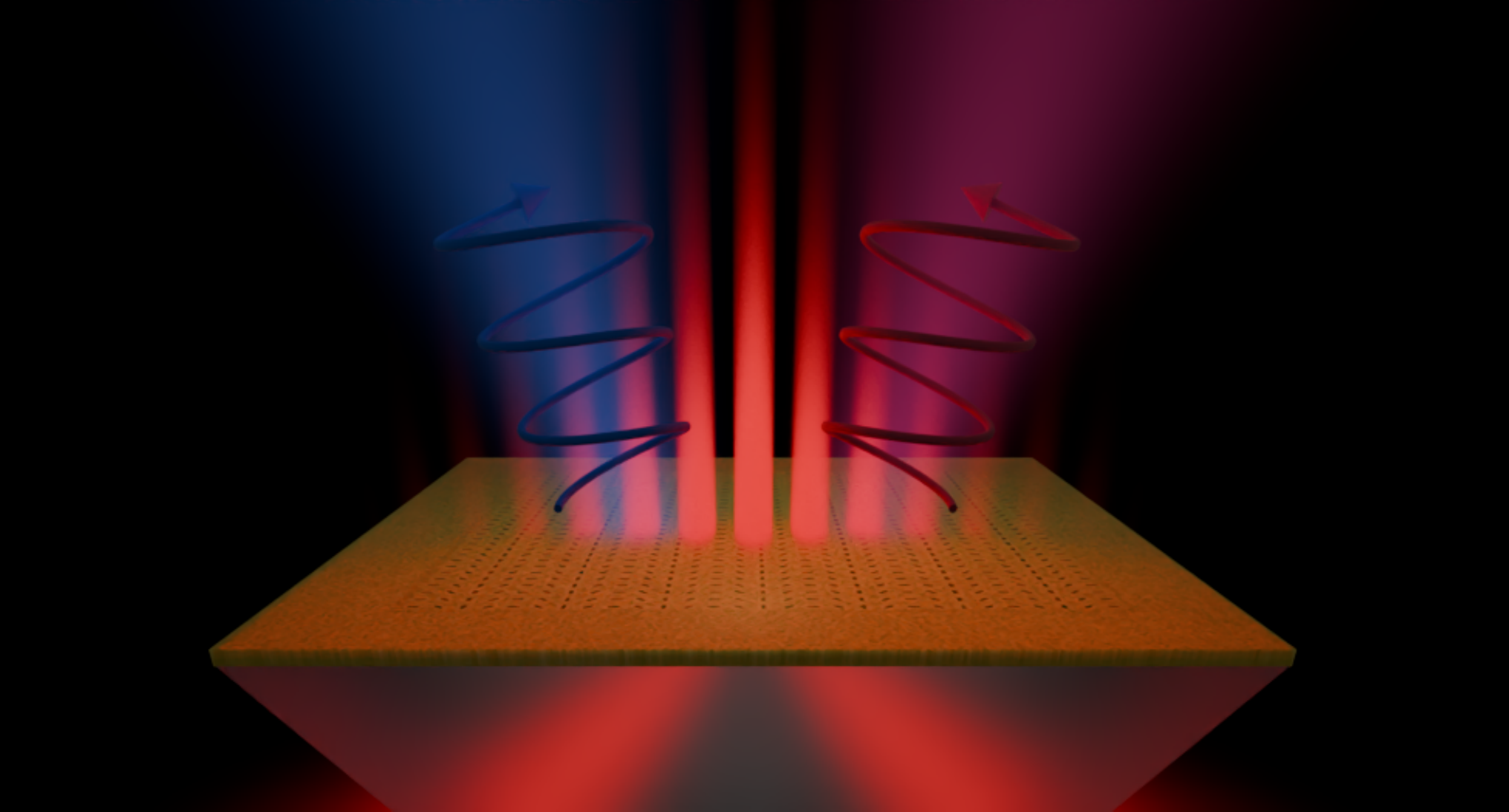}
\end{tocentry}

\begin{abstract} 
Nonreciprocity is an important scientific concept related to the broken symmetry of light propagation through a system in forward and reverse directions. 
This effect lies in the origin of various applications including signal processing, noise reduction, unidirectional propagation and sensing.
Here we show that propagation of Surface Plasmons (SP) within a structure having a false chirality exhibits a non-reciprocity.
The SP waves propagating in opposite directions within the structure acquire opposite Pancharatnam-Berry (PB) phases.
To detect this phase difference we introduce a novel interferometric technique based on a customized Sagnac set-up.
The main advantages of our proposed system are high sensitivity to non-reciprocal phase changes, high precision incidence angle alignment and the inspection of the $k$-space enabled by sufficiently wide range of incidence angles.
We believe that a pivotal role of the non-reciprocity and its detection in numerous physical and chemical processes suggests a wide range of practical applications as well as deeper scientific insights.
\end{abstract}

\section{Introduction}
The discovery of polarization rotation of light passing through glass in the direction of an applied magnetic field by Faraday \cite{faraday1933faraday4} in 1845 marks the beginning of nonreciprocity studies in optics \cite{caloz2018electromagnetic}. Reciprocity (nonreciprocity) is the property of a system, in which the light propagating forward and its reverse ray have the same (different) optical adventures \cite{caloz2018electromagnetic}. The nonreciprocity has its foundation in various concepts including time-reversal (TR) symmetry breaking, Onsager-Casimir relations \cite{onsager1931reciprocal,onsager1931reciprocal2,casimir1945onsager} and the Lorentz reciprocity theorem \cite{caloz2018electromagnetic, caloz2018nonreciprocity, asadchy2020tutorial}.

Maxwell's equations and their solutions are invariant under TR \cite{altman2011reciprocity}. However, the involved physical quantities may be characterized by their parity: the electric charge, field, polarization, and displacement have even parity, while velocity, displacement current, magnetic field, wavevector, Poynting vector, gain, and loss present odd parity \cite{jackson2021classical}. TR symmetry requires that even quantities must be maintained under TR, while odd quantities change their sign. However, the system exhibits non-reciprocity between the forward and reverse propagation of light, only when the TR symmetry is broken. Practically, this can be implemented by developing a system whose response to at least one odd-parity quantity is unchanged under TR \cite{caloz2018electromagnetic, caloz2018nonreciprocity, asadchy2020tutorial}.

To date, non-reciprocity has been observed in a variety of phenomena that include, among others, the Sagnac effect \cite{sagnac1914effet}, Fresnel drag effect \cite{huidobro2019fresnel}, Faraday effect \cite{faraday1933faraday4}, and Kerr polarization effect \cite{kaplan1981enhancement, kaplan1982directionally, kaplan1983light}. It is essential for applications where unidirectional propagation is required, such as radars using a single antenna for transmitting and receiving, suppression of destabilizing reflections in lasers, isolation of signals from a power supply, waveguide phase shifters, etc. \cite{asadchy2020tutorial,pozar2021microwave}

Non-reciprocity can be produced by exploiting the concept of chirality, which is the property of a system that can exist in two distinct enantiomeric states of opposite handedness. A system can be classified as possessing a true or a false chirality. An object/influence having a true chirality can be brought to coincide with its mirror image only by space inversion and not by any combination of time inversion and proper spatial rotations \cite{barron1986true}. A true chiral system (e.g: Chiral molecules featuring natural optical activity) maintains its handedness under TR as opposed, for instance, to a Faraday medium \cite{faraday1933faraday4, caloz2018electromagnetic}. The latter effect results from the time-odd circular birefringence arising from a false-chiral spin states arrangement in the medium due to an external magnetic field  \cite{barron1986true2, barron1986symmetry, de1980pictorial, caloz2018electromagnetic}. Consequently, one can conclude that a system possessing a false chirality is nonreciprocal \cite{ bliokh2014magnetoelectric}.

%In this letter, we observe a non-reciprocity induced by the Pancharatnam-Berry phase in a plasmonic metasurface featuring false chirality.
In this letter, 
%we show that a plasmonic metasurface featuring \textit{false chirality} exhibits non-reciprocity due to a Pancharatnam-Berry phase.
\textcolor{black}{we demonstrate, an innovative approach for achieving and directly measuring nonreciprocity in plasmonic systems by implementing false chirality in a PB metasurface. Unlike other nonreciprocal plasmonic systems that rely on electric \cite{hassani2022drifting} or magnetic \cite{hassani2019unidirectional} biasing to induce asymmetry, our system exhibits nonreciprocity inherently due to false chirality, without the need for external fields, biases or non-linearity \cite{guo2022nonreciprocal}.}  
In our system, surface plasmons -- collective oscillations of bound charge densities -- are excited in the vicinity of a metal surface by using a Kretschmann's configuration.
These surface waves propagate on a metasurface comprised of spatially rotated rectangular apertures. The proposed 2D metasurface can comprise two distinct enantiomers (depending on rotation handedness) but can be interconverted by a proper spatial rotation, therefore it is classified as false chiral.
The plasmonic signal interacting with the structure radiates circularly polarized photons into free space.
The momentum matching in this configuration is modified by the Pancharatnam-Berry phase due to the periodic rotation of the polarization state in the near-field \cite{pancharatnam1956generalized1, pancharatnam1956generalized,berry1987adiabatic,jisha2021geometric,cohen2019geometric,samuel1988general,de2012pancharatnam}, rigorously demonstrated by a microscopic analysis for equivalent metasurfaces \cite{loren2023microscopic}.
This breaks the symmetry of the counter-propagating plasmonic modes in the $k$-space leading to a non-reciprocal phase lag.
\textcolor{black}{Previously, we have shown that similar metasurface can selectively excite plasmonic modes depending on the state of the incident polarization \cite{fox2022generalized}, which points to the potential nonreciprocity of the metasurface. Here we use a customized} Sagnac interferometer \cite{culshaw2005optical} to detect this phase in a fringe pattern  as compared to an unpatterned gold surface.
We verify that non-chiral arrays do not affect systems reciprocity and behave similarly to the plain surface.
Measurements were performed with different geometric parameters of the gratings and the excitation, proving a well pronounced sensitivity to the local dielectric properties and the structural chirality.
This enables the potential implementation of our system as an instrument for sensing nonreciprocal analytes including enantiomeric biomolecules \cite{homola2003present} and nanostructures.
This may significantly contribute to the development of ultrasensitive bio-photonic devices and lab-on-chip systems.

\section{Experimental Results and Discussion}

\subsection{SP-based Sagnac Interferometer}

\begin{figure}[htbp]
    \centering
    \includegraphics[width=1\textwidth]{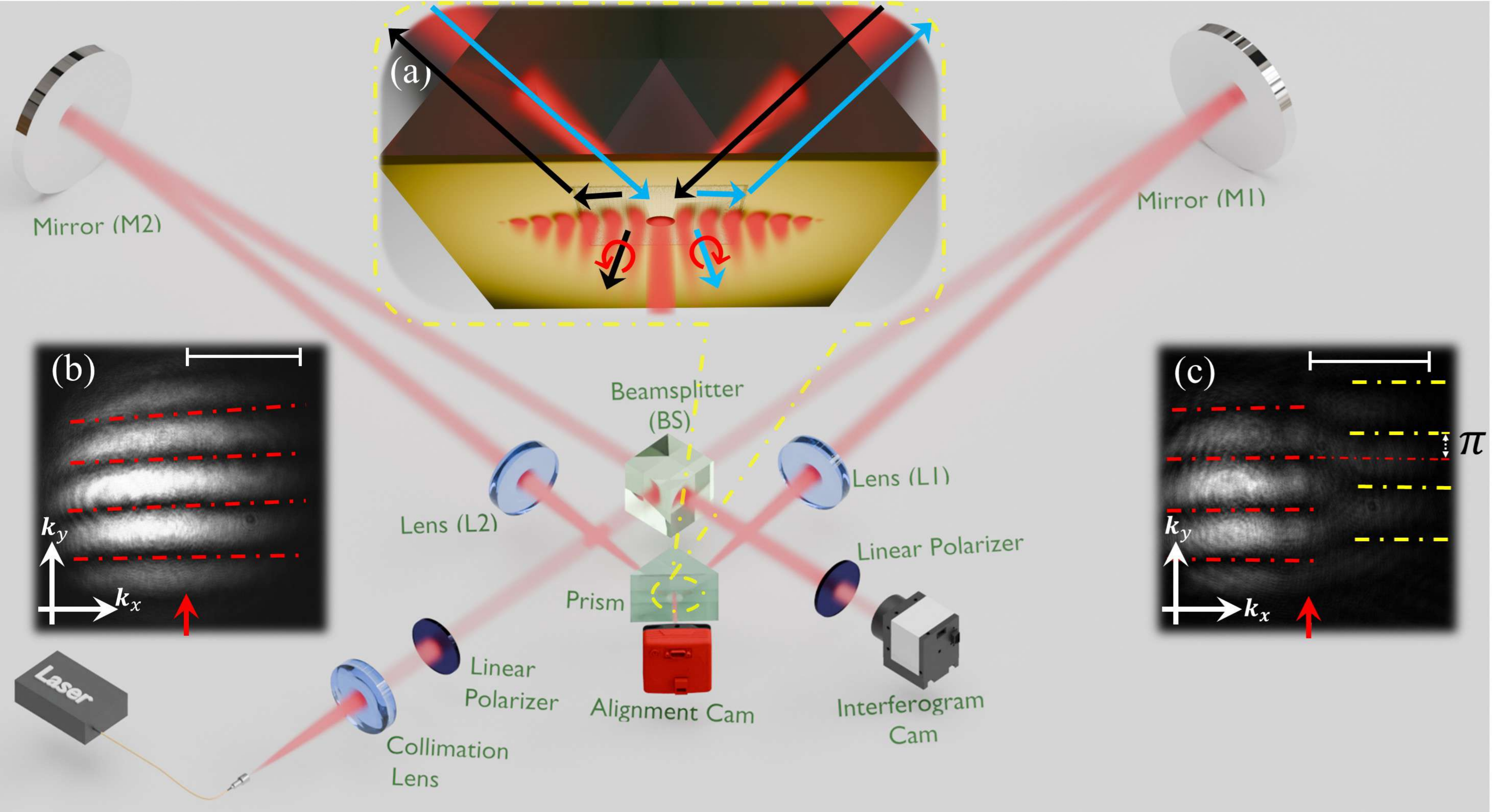}
    \caption{Sagnac Interferometer Setup (see details in the text).
    Inset (a) illustrates SPP excitation in the Kretschmann configuration.
    The black and blue arrows represent the CW and CCW paths, respectively.
    Inset (b) \& (c) depict the obtained interferograms for non-TM and TM polarized incident beams. The red arrow denotes the angle/momentum for SP resonance. The scale spans $0.48/\mu m$ in $k$-space ($\sim \, 3.02^{o}$ angle in real space).}
    \label{fig:Sagnac Setup}
\end{figure}

The scheme of the experimental setup is depicted in Fig. \ref{fig:Sagnac Setup}.
A diode laser (LP785-SAV50, $\lambda_0 = 785$ nm) was initially expanded and collimated using a combination of lenses.
The laser beam then passed through a linear polarizer (LP) to ensure the required TM polarization state, verified using a digital polarimeter. %\cite{caucheteur2015review, berini2009long}

Subsequently, the beam was sent into the Sagnac loop, which was the core part of the setup, comprised of a non-polarizing beamsplitter (BS), two mirrors (M1 \& M2), two focusing lenses ($L1$ \&$L2$) and the prism-sample assembly.
The samples were fabricated using a focused ion beam (FIB, dual-beam Helios 5) etching inside a 75 nm gold film evaporated on a 160 $\mu m$ thick cover slip.
The structures comprised periodic arrays of apertures etched through the gold film. 
The plasmonic sample with its substrate was attached to the back side of a prism using index-matching oil, thus providing the necessary geometry for the SP excitation on the metal by means of the Krestchmann configuration.
Due to the BS, the plasmonic wave was excited by two counter-propagating TM-polarized beams (as shown in the inset of Fig. \ref{fig:Sagnac Setup} (a)) providing a clockwise (CW) and a counter-clockwise (CCW) paths.  
After passing again through the BS, the recombined beam was sent through a second LP, aligned perpendicularly to the initial TM state.
The resultant interferogram was then formed and recorded by the CMOS camera (DCC3240C).

The custom \textit{triangular configuration} of the Sagnac Interferometer was implemented in order to make the setup alignment easier and facilitate a precise fine-tuning of the incidence angle.
Accordingly, the plasmonic resonance angle for both beams was obtained exactly in the middle of the spot.
To obtain a base interferogram with horizontal fringes, a slight vertical tilt ($\sim 0.01\,rad$) was introduced between the interferometer arms.
The bare interference pattern arising with the TE(TM) polarized illumination at a flat surface is presented in Fig. \ref{fig:Sagnac Setup} b(c).
We note the appearance of the central dark line (along the red arrow in Fig. \ref{fig:Sagnac Setup} c) indicating the SP resonance with the TM illumination only.
The lenses $L1$ and $L2$ were used to focus an incoming quasi-plane-wave beam into a small spot on the interface and revert the reflected part to a plane-wave, so the images received in the interferogram camera could be considered as a $k$-space, representing the angular spectrum of the incident light.
A vertical fringe shift of $\pi$ was observed in the interferogram across the SP resonance.
This is a feature of the system arising due to the usage of two polarizers that are orthogonal to each other.
The excitation of SPs on a flat gold surface was expected to maintain the system reciprocity, since the phases acquired in both arms were identical.
Accordingly, this measurement was used as a base interferogram for the detection of the non-reciprocity of system leading to an additional fringe shifting.

\subsection{Direct Measurement of a Structure Non-reciprocity}

\begin{figure}[htbp]
    \centering
    \includegraphics[width=1\linewidth]{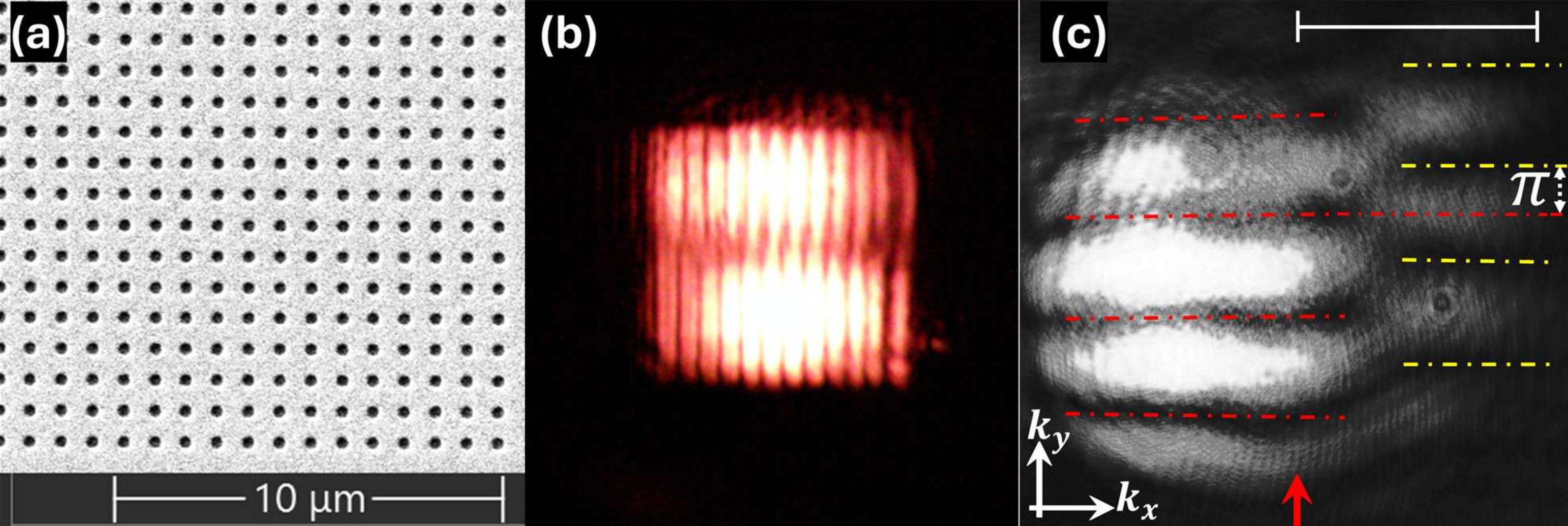}
    \caption{The Hole Array experiment (Exp HA) featuring reciprocal light-matter interaction.
    (a) SEM image of isotropic Hole Array grating utilized in Exp HA.
    (b) Microscope image of Exp HA showing light emission under momentum matching when the TM polarized incident beams are focused on it.
    (c) The interferogram obtained for Exp HA. The red arrow denotes the angle/momentum for SP resonance. The scale spans $0.48/\mu m$ in $k$-space ($\sim \, 3.02^{o}$ angle in real space).}
    \label{fig: result1}
\end{figure}

First we studied the interaction of SPs with a non-chiral structure hypothesizing that such a geometry did not break the TR symmetry.
We used a simple square array of circular holes of 300 nm in diameter (as shown in Fig. \ref{fig: result1} (a)) with a period of $\Lambda = 795$ nm.
The SPs propagating through this grating encounter periodic scatterers leading to the light out-coupling to free space.
The distribution of the emitted radiation in the upper hemisphere (above the prism) is conveniently described by the momentum equation in 2D: $\mathbf{k}_{out} = \mathbf{k}_{SP}\pm \frac{2m\pi}{\Lambda} \mathbf{\hat{x}}\pm \frac{2n\pi}{\Lambda} \mathbf{\hat{y}}$, with $m$ and $n$ as integers and the SP wave number given by the dispersion relation,  $k_{SP}=\frac{2\pi}{\lambda_0}\sqrt{\frac{\varepsilon_{m}\varepsilon _{d}}{\varepsilon_{m}+\varepsilon_{d}}}$, where $\varepsilon_{m/d}$ stands for the dielectric constants of the gold and the air, respectively.
The light emitted from the array to the free space was captured by the alignment camera and shown in Fig. \ref{fig: result1} (b).
We note that due to the requirement of the relative tilt between the interfering beams, the CW and the CCW paths generate two slightly displaced spots on the sample.
We carefully aligned the optics to ensure that both spots appear within the sample area to retain the balance between the arms.

When observing the image obtained with the interferogram camera for the hole array (see Fig. \ref{fig: result1} (c)) we find that the fringe pattern is similar to the base interferogram with a characteristic phase shift of $\pi$ around the SP resonance line.
This indicates that the phases acquired in two counter-propagating beams are identical and the system maintains the reciprocity as has been suggested from the grating symmetry.

\begin{figure}[htbp]
    \centering
    \includegraphics[width=0.8\linewidth]{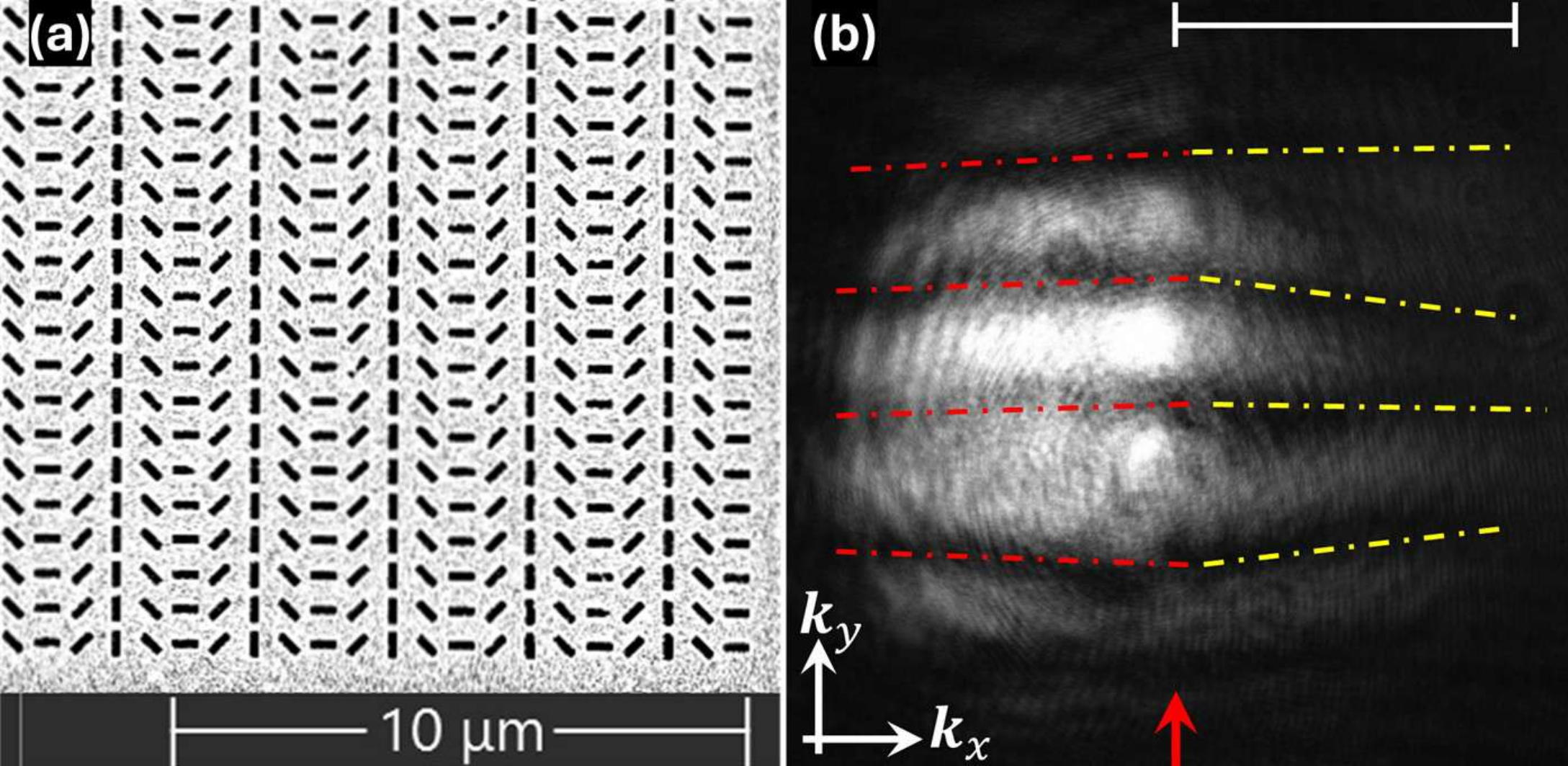}
    \caption{False chiral grating experiment. (a) The SEM image of the ShP - Shorter period grating (RH, 596.75 nm). (b) The interferogram from the grating. The red arrow denotes the angle/momentum for SP resonance. The scale spans $0.48/\mu m$ in $k$-space ($\sim \, 3.02^{o}$ angle in real space).}
    \label{fig: result2}
\end{figure}

In order to break the symmetry we used plasmonic structures with the built-in false chirality.
The structure comprised of rectangular (150 x 450 nm) apertures whose orientation $\theta(x)$ was varied along the $x$ axis with the rate of $\Omega_x = \dot{\theta}(x) = \pi/N \Lambda$ where $\Lambda$ is the grating period and $N$ is the number of periods needed to rotate the aperture by $\pi$ radians.
The precise choice of these latter parameters have been done in accordance with the desired momentum matching condition and will be discussed later.

A typical grating with spatially rotated rectangular apertures is shown in Fig. \ref{fig: result2} (a).
Clearly, the SP wave propagating through this structure is characterized by a false chirality, since under TR the plasmon encounters an opposite rotation handedness.
Structures with a shorter-period (ShP) of $\Lambda = 597$ nm (see Fig. \ref{fig: result2} (a)) and with a longer period (LoP) of $\Lambda = 995$ nm (see Supporting Information below) were fabricated.
Figure \ref{fig: result2} (b) summarizes the interferometric measurement of the ShP grating structure.
It is noteworthy that the result shows a clear phase shift with respect to the a-priori measured base interferogram (see Fig. \ref{fig:Sagnac Setup} (c)).
Depending on the period and the structure handedness the relative fringe shift between the left and the right part of the spot will decrease or increase by $\pi$ radians; causing the disappearance of the phase dislocation from Fig. \ref{fig: result2} (b).
This result confirms that the plasmonic waves propagating through our structure experience a time-reversal symmetry breaking leading to the system non-reciprocity.
However, to understand this phenomenon better, one must delve into a physical mechanism of SP propagation within the structure with false chirality. 

A propagating plasmonic wave encountering a rectangular aperture excites a dipole-like scattering aligned with the aperture orientation $\theta(x)$.
The rotation of the apertures along $x$ direction leads to a periodically rotated linear polarization state of the emitted light.
Metasurfaces with hexagonal and square array rectangular apertures rotated along a single or multiple axes have been experimentally studied \cite{fox2022generalized, fox2023topologically} and analytically modeled \cite{loren2023microscopic} for discrete systems.
It was shown that the resulting modified momentum equation yields,

\begin{equation}
\mathbf{k}_{out} = \mathbf{k}_{SP}\pm \frac{2m\pi}{\Lambda} \mathbf{\hat{x}}\pm \frac{2n\pi}{\Lambda} \mathbf{\hat{y}} - 2\sigma\Omega\mathbf{\hat{x}},
\label{eqn: momentum}
\end{equation}
where $\sigma = \pm1$ represents the handedness of the emitted circular state \cite{fox2022generalized}.
In general $\sigma = +1$ ($\sigma = -1$) corresponds to the right-handed (left-handed) circular polarization (RCP/LCP) while $\sigma = 0$ stands for the linear TM state (similar to the hole-array case).
The last term in the Eq. \ref{eqn: momentum} represents the polarization dependent Pancharatnam-Berry phase \cite{fox2022generalized}, $\varphi_{PB} = 2\sigma \theta (x)$.

\begin{figure}[htbp]
    \centering
    \includegraphics[width=1\linewidth]{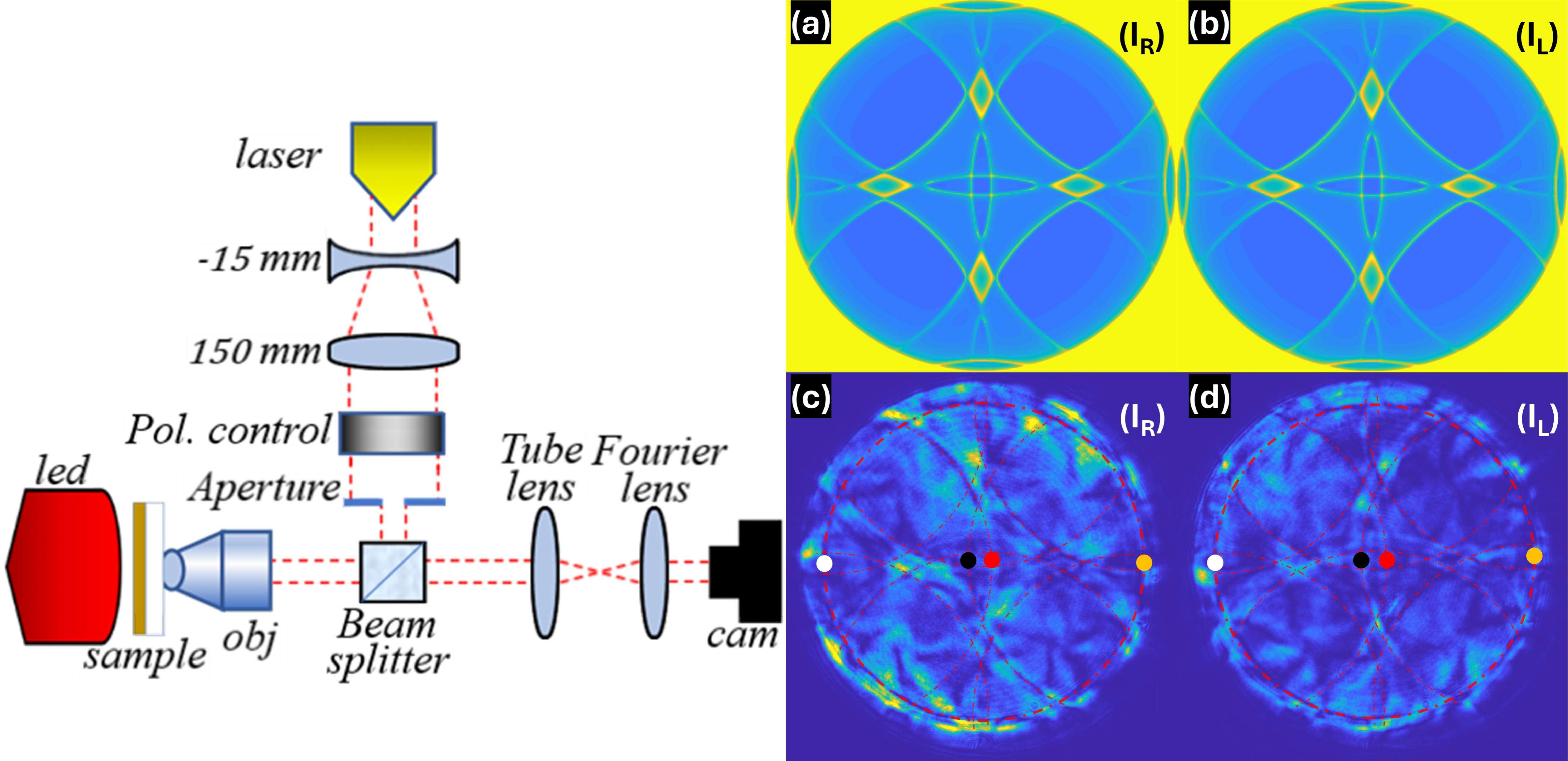}
    \caption{The inverted leakage-radiation microscopy (LRM) setup and the k-space representation of SP dispersion modes for the non-chiral metasurface.
    The simulated $k$-space result for the square Hole Array Experiment with: (a) RCP \& (b) LCP excitation of SPs.
    The inverted LRM measured $k$-space results for: (c) RCP \& (d) LCP excitations.
    In (c) \& (d), the thicker red dashed circle of radius \(k_{SP}\) marks the positions of primary SP resonance mode while other red dashed circles represent its Bloch replicas.
    The white and orange spots represent the primary SP resonance mode for SPs strictly propagating in the $-\hat{x}$ and $+\hat{x}$ directions respectively.
    The black and red spots are their corresponding Bloch replicas.}
    \label{fig: ILRM_HA}
\end{figure}

To obtain the full $k$-space of our structure we used an \textit{inverted} leakage-radiation microscopy (LRM) system with a Fourier lens (depicted in Fig. \ref{fig: ILRM_HA}) before the camera.
The input circular polarization was controlled by a quarter-wave plate (QWP) placed before the sample.
We started by studying the non-chiral structure - a square hole array previously investigated in Fig. \ref{fig: result1}.
The intensity distribution in the $k$-space was separately measured with the RCP and the LCP incident states ($I_R$ and $I_L$) as can be seen in Fig. \ref{fig: ILRM_HA} (c, d), respectively.
The dashed red circles depict the locations of the primary SP resonance and its Bloch replications due to  $\pm \frac{2m\pi}{\Lambda}$ and $\pm \frac{2n\pi}{\Lambda}$ terms in Eq. \ref{eqn: momentum}.
We also rigorously calculated these plasmonic modes.
These calculations are based on the coupled-mode method or eigenmode expansion, which is a linear frequency-domain method that relies on the decomposition of the electromagnetic fields into a basis set of local eigenmodes of the simulated media, found by solving Maxwell's equations \cite{loren2023microscopic, loren2023spinmomentum} and presented the results in Fig. \ref{fig: ILRM_HA} (a, b).
The circular arcs represent the plasmonic modes/loss channels in the momentum space.
For instance, when a plasmon propagates strictly along a positive $x$ direction (depicted by an orange spot on the primary resonance line), it is partially converted to a free space radiation at an angle represented by the corresponding point (red spot) on the replicated resonance line.
The period of the array was chosen such that this point appears only once within the radius defined by $NA = 1$.
The negative propagation of the SP is represented by the white spot and results in the emission from the point marked by the black spot.
The separation between the black and the red spot means that the counter-propagating SP waves are not coupled.
It is evident that besides small intensity fluctuations, both the experimental and the calculated results are very similar for opposite polarizations.
In particular, one may note that both polarizations result in the same amount of plasmonic modes.
Since the losses are equal for the plasmons propagating in opposite directions and independently of the polarization, the system stays reciprocal as was shown by the interference experiment.

\begin{figure}[htbp]
    \centering
    \includegraphics[width=0.5\linewidth]{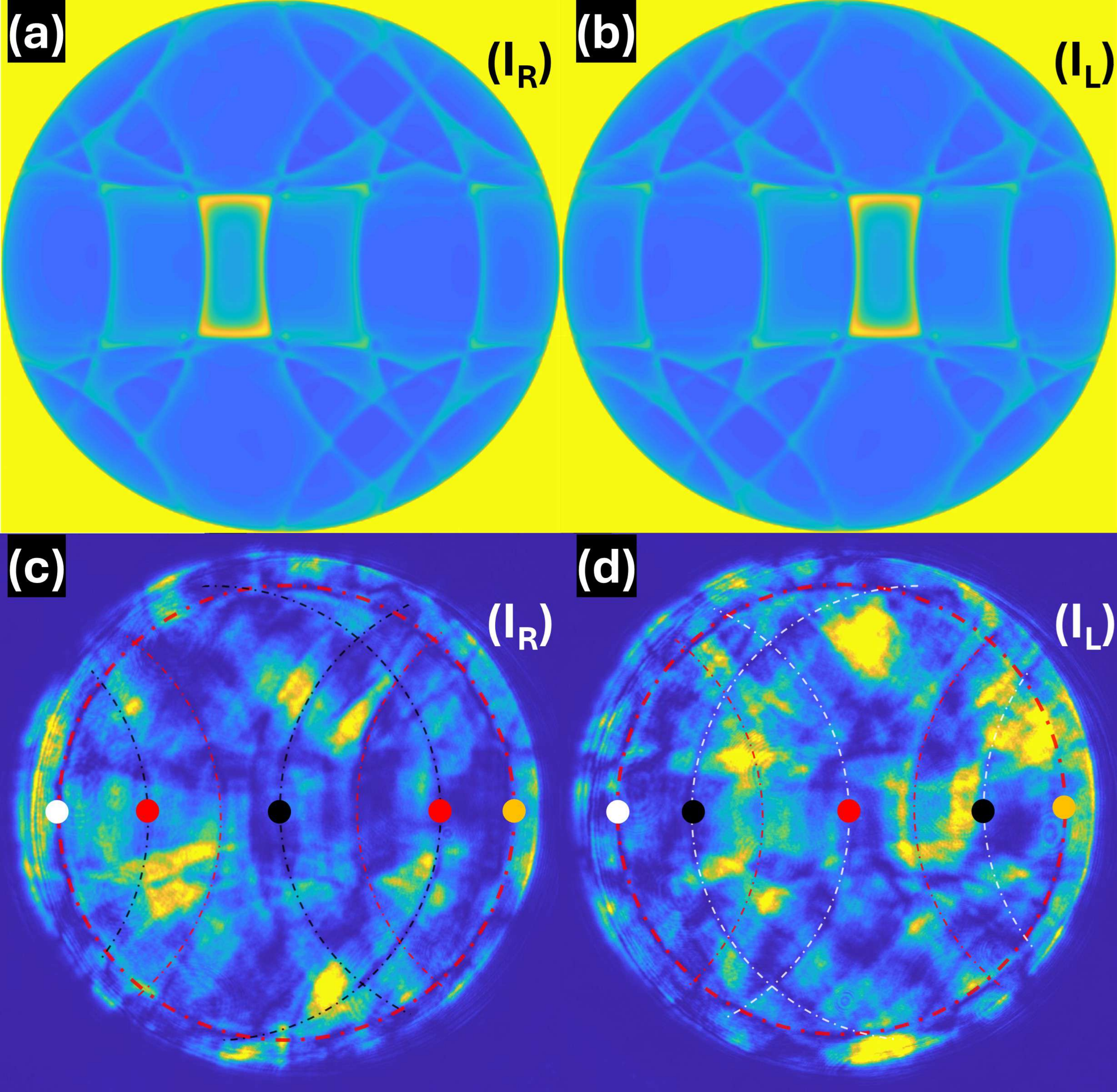}
    \caption{The k-space representation of SP dispersion modes for the false-chiral metasurface with Shorter Period (ShP) gratings.
    The simulated $k$-space result for: (a) RCP \& (b) LCP excitation of SPs.
    The inverted LRM measured $k$-space results for: (c) RCP \& (d) LCP excitation of SPs.
    In Figs. (c) \& (d), the thicker red dashed circle of radius \(k_{SP}\) marks the position of primary SP resonance mode while other red dashed circles represent its Bloch replicas.
    The black and white dashed circles respectively denote the spin-dependent SP resonance modes with right and left-handedness.
    The white and orange spots represent the primary SP resonance mode for SPs strictly propagating in the $-\hat{x}$ and $+\hat{x}$ directions respectively.
    The black and red spots are the spin dependent SP resonance modes strictly propagating in the $-\hat{x}$ and $+\hat{x}$ directions respectively.}
    \label{fig: ILRM_ShP}
\end{figure}

\begin{figure}[htbp]
  \centering
  \includegraphics[width=0.6\linewidth]{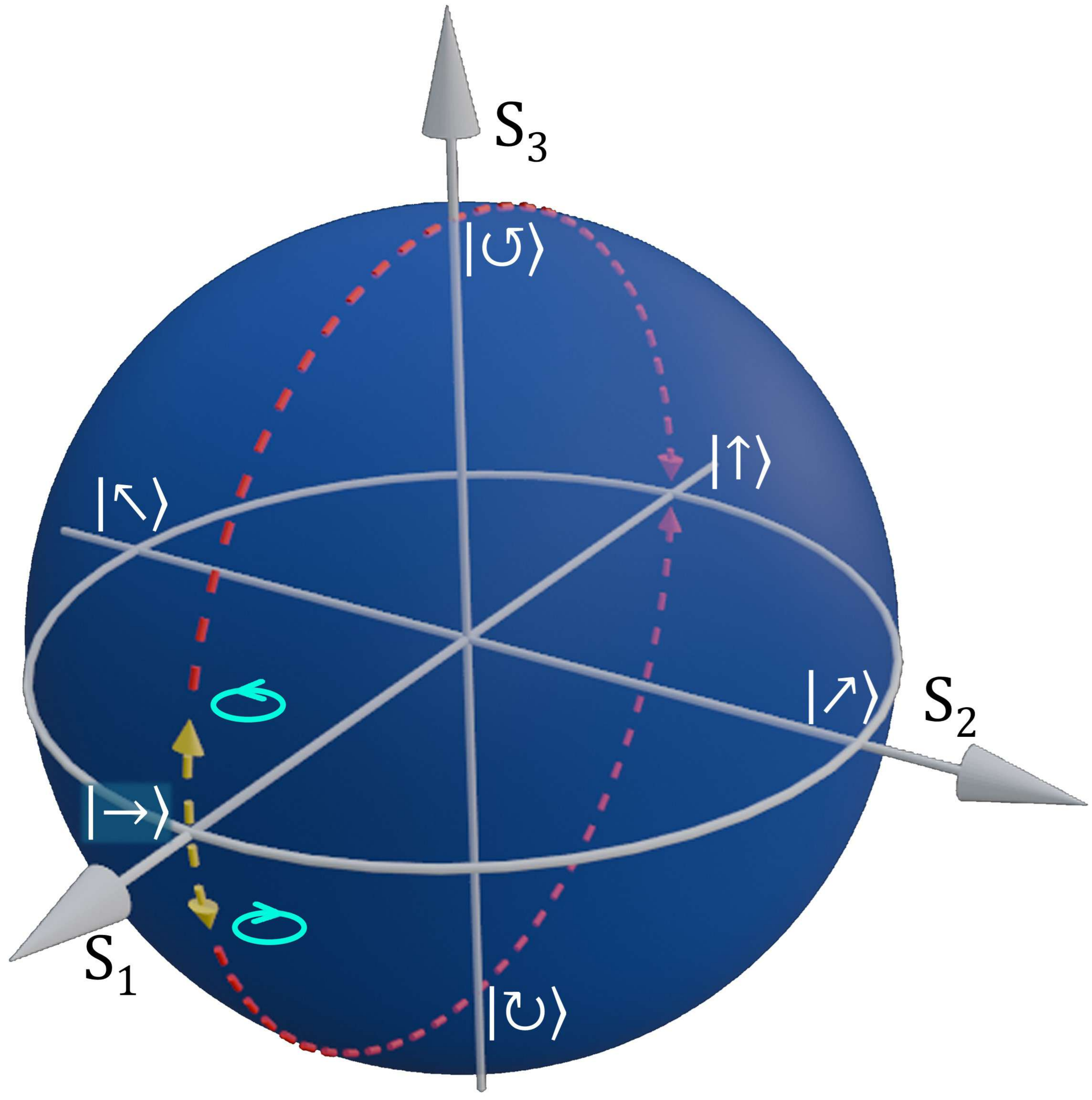}
  \caption{Poincar\'{e} sphere representation of the transport of the polarization state by the nonreciprocal system. The yellow dashed arrows represent the polarization state changes in CW and CCW beams due to the false chiral metasurface. The red dashed arrows represent the collapse of the polarization states by the TE analyzer to the $S_{1}=-1$ state.}
  \label{fig: Poincare}
\end{figure}

The situation changes for the false chiral structure. 
When comparing the measurements for the RCP and the LCP incident light we notice a striking difference in the resonance lines arrangement.
In the simulated and the experimental results (see Fig. \ref{fig: ILRM_ShP}), one may recognize modes that appear in both spin-states (depicted by the red dashed lines) as opposed to polarization-dependent modes marked by the black or the white dashed lines.
These modes are represented by the last term in Eq. \ref{eqn: momentum} with an appropriate choice of $\sigma$.
Following the same logic from the HA experiment, we find that a right-propagating SP (marked by an orange spot) has two emission angles at the RCP state (two red spots in Fig. \ref{fig: ILRM_ShP} (c)) and only one angle of emission at the LCP state (one red spot in Fig. \ref{fig: ILRM_ShP} (d)).
From this experiment one can conclude that counter-propagating SPs in the $x$ direction experience a different polarization modulation due to an unbalanced emission of circularly polarized light.
As a result, the light reflected back to the system becomes elliptically polarized with a handedness depending on the direction of propagation.
The results exhibit additional plasmonic modes due to the $y$ periodicity, however, the polarization dependence vanishes along $k_y$ axis as expected from the symmetry of the structure.

The opposite handedness of the counter propagating beams reflected from a rotated apertures' array can be conveniently modeled by direction-dependent dichroic Jones matrices, $\mathbf{R}_{CW} = \begin{pmatrix} 1&0\\0&1-\delta\\ \end{pmatrix}$ and $ \mathbf{R}_{CCW} = \begin{pmatrix} 1-\delta&0\\0&1\\ \end{pmatrix}$. These matrices are represented in the basis of circular polarizations, where 
$\left |\circlearrowright\right > = [1,0]^T$ and $\left |\circlearrowleft\right > = [0,1]^T$ are RCP and LCP states, respectively. \textcolor{black}{We assume for the sake of the model that the losses do not reduce the degree of polarization.} Accordingly, the field \textcolor{black}{loss due to the differential emission} $1>\delta>0$ is selectively applied to a corresponding circular state depending on the propagation direction. 
The TM and TE polarized light, represented as
$ \left | \rightarrow \right > = [1,0]^T$ and $\left | \uparrow \right > = [0, 1]^T$
can be translated into the helical basis by means of the matrix 
$\mathbf{\tilde{T}} = \frac{1}{\sqrt{2}}\begin{pmatrix} 1&i\\1&-i\\ \end{pmatrix}$ \cite{bomzon2001pancharatnam}.
Therefore the effect of the light passing a single loop in our Sagnac interferometer is described by the following sequence,
\begin{equation}
\mathbf{P}_{y} \mathbf{\tilde{T}}^\dagger \mathbf{R}_{cw/ccw} \mathbf{\tilde{T}} \left | \rightarrow \right > = i (\delta/2) \left | \uparrow \right > e^{i\phi_{cw/ccw}} 
\label{eqn: Jones}
\end{equation}
where the matrix $\mathbf{P}_{y} = \begin{pmatrix} 0&0\\0&1\\ \end{pmatrix}$ represents the operation of the TE analyzer placed before the camera. This calculation yields that even for a tiny RCP/LCP \textcolor{black}{loss}, a constant phase difference of $\pi$ arises between the CW and the CCW beams. This resembles the effect of the ``weak measurement'' in a plasmonic nanoslit \textcolor{black}{where a weak deviation of the incident polarization state from being parallel to the slit leads to a giant plasmonic beam shifts.} \cite{gorodetski2012weak}.
Additionally, we have confirmed this simple analytical model with numerical simulations, obtaining a sign change in the TE reflection coefficient depending on the propagation direction.

Alternatively, our system can be described as follows:
The evolution of the light polarization state in the system can be depicted on the surface of the Poincar\'{e} sphere \cite{collett1992polarized} (see Fig. \ref{fig: Poincare}).
The incident TM polarization ($S_1 = 1$) interacts with the structure and becomes elliptical as marked by the two yellow dashed arrows moving up and down along the geodesic line.
Before the interferogram camera, both beams are passing through a TE analyzer leading to the collapse %(marked by the red dashed arrows)
of the polarization states to the $S_1=-1$ point, subsequently resulting in a geometric phase of $\pi$, also corresponding to the half of the area enclosed on the sphere.
When the resulting fringe pattern is compared to that from the flat surface, a $\pi$ phase shift is clearly visible.
Altering the structure periods modifies the momentum equation (Eq. \ref{eqn: momentum}) and a different number of polarization-dependent plasmonic modes can appear within the emission hemisphere ($NA=1$).
The density of these modes and the number of line crossings (inter-mode couplings) may affect the polarization effect of the system, however, the false-chirality of the structure always results in a non-reciprocity.

\begin{figure}[htbp]
  \centering
  \includegraphics[width=1\linewidth]{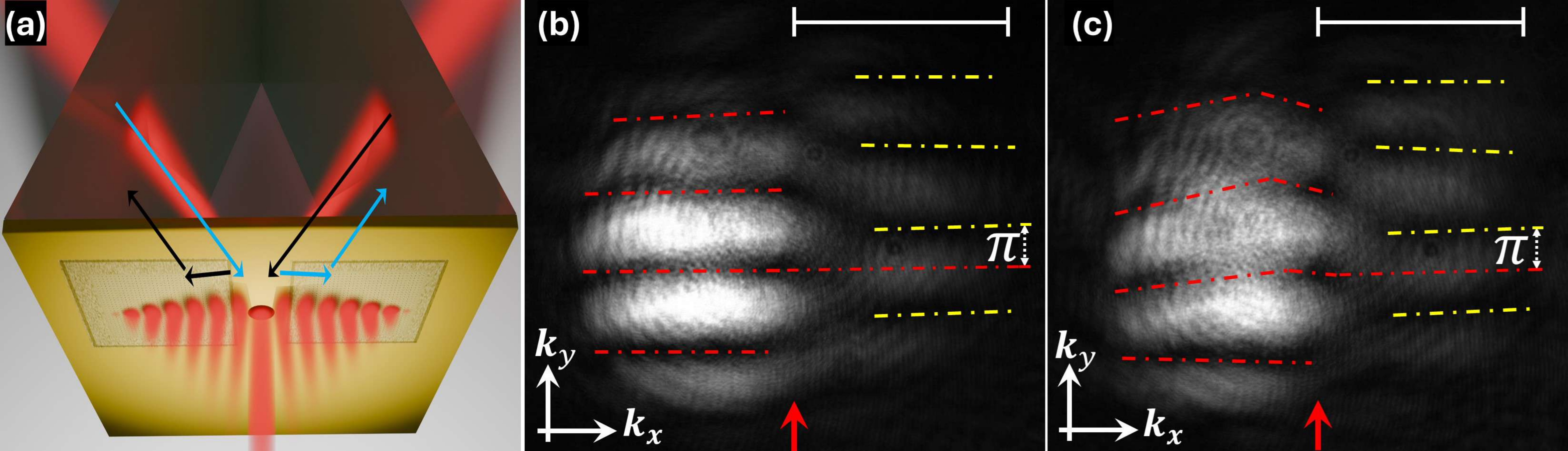}
  \caption{Experiment to study the effect of the beam incidence’s proximity to the metasurface.
  (a) The experimental scheme with two square regions of gratings.
  Interferograms obtained from two false chiral LoP grating squares with a centre-to-centre separation of (b) \(70 \mu m\) and (c) \(40 \mu m\). The red arrow denotes the angle/momentum for SP resonance. The scale spans $0.48/\mu m$ in $k$-space ($\sim \, 3.02^{o}$ angle in real space).}
  \label{fig: result3}
\end{figure}

Finally we examined the effect of the distance between the excitation spot and the sample.
In all the previous experiments we focused the incident light at sample centers.
Now we fabricated pairs of similar samples (False-Chiral metasurface with LoP structure) separated by a distance $d = 70\mu m, 40 \mu m$.
The beam was then centered between the structures so that the excited SP wave had to propagate half of the distance $d$ before encountering the gratings.
Figure \ref {fig: result3} shows the measured interferograms.
It is clearly visible that the interferograms are similar to the reciprocal case and the additional $\pi$ phase related to the nonreciprocity vanishes for both cases which might be an evidence of a very weak interaction with the structures.
We suppose that the reason for that might be additional losses in SP propagation and coupling to light.

\subsection{Summary}
\textcolor{black}{In summary, we have experimentally demonstrated that a plasmonic PB metasurface featuring  false chirality induces non-reciprocity which was measured  using a custom Sagnac interferometer setup.} 
%In summary, we conducted a direct far field phase measurement of the non-reciprocity, induced by a plasmonic metasurface featuring false chirality, using a custom Sagnac interferometer setup.
The false chiral nature of the grating metasurfaces results in the transport of the polarization state on the Poincar\'{e} surface and introduces non-reciprocity (attributed to the accrued Pancharatnam-Berry phases).
This leads to a differential spin dependent absorption in counter propagating SPs and hence introduces a measurable phase shift that was measured by an interferometric method.
A simulation and an inverted leakage radiation microscopy experiment was performed to study the plasmonic mode dispersion of SPs excited on the metasurface.
Both the simulation and the experiment consistently showed that the non-reciprocity in our system is related to the spin-dependent asymmetric plasmonic mode dispersion.
We tested structures with a different handedness and a period and also a non-chiral structure to verify the robustness of our measurements.
We believe that the custom implementation of the Sagnac interferometer demonstrated here can lead to highly sensitive phase interrogation measurement related to surface plasmon resonances and to measure the non-reciprocity introduced by subwavelength analytes and nanostructures.
Also, the high sensitivity of the non-reciprocal effect to the excitation of SPs on the metasurface suggest that our system can be utilized for device miniaturization and lab-on-chip systems.

\section{Data Availability Statement}
The data that support the findings within this paper are available from the corresponding authors upon request. 

%\begin{acknowledgement}

%Please use ``The authors thank \ldots'' rather than ``The
%authors would like to thank \ldots''.

%\end{acknowledgement}

\begin{suppinfo}
Additional experimental data with larger period metasurface.
\end{suppinfo}

%\bibliography{Ref}

\providecommand{\latin}[1]{#1}
\makeatletter
\providecommand{\doi}
  {\begingroup\let\do\@makeother\dospecials
  \catcode`\{=1 \catcode`\}=2 \doi@aux}
\providecommand{\doi@aux}[1]{\endgroup\texttt{#1}}
\makeatother
\providecommand*\mcitethebibliography{\thebibliography}
\csname @ifundefined\endcsname{endmcitethebibliography}
  {\let\endmcitethebibliography\endthebibliography}{}
\begin{mcitethebibliography}{40}
\providecommand*\natexlab[1]{#1}
\providecommand*\mciteSetBstSublistMode[1]{}
\providecommand*\mciteSetBstMaxWidthForm[2]{}
\providecommand*\mciteBstWouldAddEndPuncttrue
  {\def\EndOfBibitem{\unskip.}}
\providecommand*\mciteBstWouldAddEndPunctfalse
  {\let\EndOfBibitem\relax}
\providecommand*\mciteSetBstMidEndSepPunct[3]{}
\providecommand*\mciteSetBstSublistLabelBeginEnd[3]{}
\providecommand*\EndOfBibitem{}
\mciteSetBstSublistMode{f}
\mciteSetBstMaxWidthForm{subitem}{(\alph{mcitesubitemcount})}
\mciteSetBstSublistLabelBeginEnd
  {\mcitemaxwidthsubitemform\space}
  {\relax}
  {\relax}

\bibitem[Faraday(1933)]{faraday1933faraday4}
Faraday,~M. \emph{Faraday’s diary: Being the various philosophical notes of
  experimental investigation, Vol. IV, Nov. 12, 1839 - June 26, 1847}; George
  Bell and Sons, Ltd: London, 1933\relax
\mciteBstWouldAddEndPuncttrue
\mciteSetBstMidEndSepPunct{\mcitedefaultmidpunct}
{\mcitedefaultendpunct}{\mcitedefaultseppunct}\relax
\EndOfBibitem
\bibitem[Caloz \latin{et~al.}(2018)Caloz, Alu, Tretyakov, Sounas, Achouri, and
  Deck-L{\'e}ger]{caloz2018electromagnetic}
Caloz,~C.; Alu,~A.; Tretyakov,~S.; Sounas,~D.; Achouri,~K.;
  Deck-L{\'e}ger,~Z.-L. Electromagnetic nonreciprocity. \emph{Physical Review
  Applied} \textbf{2018}, \emph{10}, 047001\relax
\mciteBstWouldAddEndPuncttrue
\mciteSetBstMidEndSepPunct{\mcitedefaultmidpunct}
{\mcitedefaultendpunct}{\mcitedefaultseppunct}\relax
\EndOfBibitem
\bibitem[Onsager(1931)]{onsager1931reciprocal}
Onsager,~L. Reciprocal relations in irreversible processes. I. \emph{Physical
  Review} \textbf{1931}, \emph{37}, 405\relax
\mciteBstWouldAddEndPuncttrue
\mciteSetBstMidEndSepPunct{\mcitedefaultmidpunct}
{\mcitedefaultendpunct}{\mcitedefaultseppunct}\relax
\EndOfBibitem
\bibitem[Onsager(1931)]{onsager1931reciprocal2}
Onsager,~L. Reciprocal relations in irreversible processes. II. \emph{Physical
  Review} \textbf{1931}, \emph{38}, 2265\relax
\mciteBstWouldAddEndPuncttrue
\mciteSetBstMidEndSepPunct{\mcitedefaultmidpunct}
{\mcitedefaultendpunct}{\mcitedefaultseppunct}\relax
\EndOfBibitem
\bibitem[Casimir(1945)]{casimir1945onsager}
Casimir,~H. B.~G. On Onsager's principle of microscopic reversibility.
  \emph{Reviews of Modern Physics} \textbf{1945}, \emph{17}, 343\relax
\mciteBstWouldAddEndPuncttrue
\mciteSetBstMidEndSepPunct{\mcitedefaultmidpunct}
{\mcitedefaultendpunct}{\mcitedefaultseppunct}\relax
\EndOfBibitem
\bibitem[Caloz \latin{et~al.}(2018, arXiv: 1804.00238 physics.optics)Caloz,
  Alù, Tretyakov, Sounas, Achouri, and Deck-Léger]{caloz2018nonreciprocity}
Caloz,~C.; Alù,~A.; Tretyakov,~S.; Sounas,~D.; Achouri,~K.; Deck-Léger,~Z.-L.
  What is Nonreciprocity? Part II. 2018, arXiv: 1804.00238 physics.optics;
  \url{https://arxiv.org/abs/1804.00238}, (accessed 2024-05-06)\relax
\mciteBstWouldAddEndPuncttrue
\mciteSetBstMidEndSepPunct{\mcitedefaultmidpunct}
{\mcitedefaultendpunct}{\mcitedefaultseppunct}\relax
\EndOfBibitem
\bibitem[Asadchy \latin{et~al.}(2020)Asadchy, Mirmoosa, Diaz-Rubio, Fan, and
  Tretyakov]{asadchy2020tutorial}
Asadchy,~V.~S.; Mirmoosa,~M.~S.; Diaz-Rubio,~A.; Fan,~S.; Tretyakov,~S.~A.
  Tutorial on electromagnetic nonreciprocity and its origins. \emph{Proceedings
  of the IEEE} \textbf{2020}, \emph{108}, 1684--1727\relax
\mciteBstWouldAddEndPuncttrue
\mciteSetBstMidEndSepPunct{\mcitedefaultmidpunct}
{\mcitedefaultendpunct}{\mcitedefaultseppunct}\relax
\EndOfBibitem
\bibitem[Altman and Suchy(2011)Altman, and Suchy]{altman2011reciprocity}
Altman,~C.; Suchy,~K. \emph{Reciprocity, spatial mapping and time reversal in
  electromagnetics}; Springer Science \& Business Media, 2011\relax
\mciteBstWouldAddEndPuncttrue
\mciteSetBstMidEndSepPunct{\mcitedefaultmidpunct}
{\mcitedefaultendpunct}{\mcitedefaultseppunct}\relax
\EndOfBibitem
\bibitem[Jackson(1998)]{jackson2021classical}
Jackson,~J.~D. \emph{Classical electrodynamics}, 3rd ed.; John Wiley \& Sons,
  1998\relax
\mciteBstWouldAddEndPuncttrue
\mciteSetBstMidEndSepPunct{\mcitedefaultmidpunct}
{\mcitedefaultendpunct}{\mcitedefaultseppunct}\relax
\EndOfBibitem
\bibitem[Sagnac(1914)]{sagnac1914effet}
Sagnac,~G. Effet tourbillonnaire optique. La circulation de l'{\'e}ther
  lumineux dans un interf{\'e}rographe tournant. \emph{J. Phys. Theor. Appl.}
  \textbf{1914}, \emph{4}, 177--195\relax
\mciteBstWouldAddEndPuncttrue
\mciteSetBstMidEndSepPunct{\mcitedefaultmidpunct}
{\mcitedefaultendpunct}{\mcitedefaultseppunct}\relax
\EndOfBibitem
\bibitem[Huidobro \latin{et~al.}(2019)Huidobro, Galiffi, Guenneau, Craster, and
  Pendry]{huidobro2019fresnel}
Huidobro,~P.~A.; Galiffi,~E.; Guenneau,~S.; Craster,~R.~V.; Pendry,~J.~B.
  Fresnel drag in space--time-modulated metamaterials. \emph{Proceedings of the
  National Academy of Sciences} \textbf{2019}, \emph{116}, 24943--24948\relax
\mciteBstWouldAddEndPuncttrue
\mciteSetBstMidEndSepPunct{\mcitedefaultmidpunct}
{\mcitedefaultendpunct}{\mcitedefaultseppunct}\relax
\EndOfBibitem
\bibitem[Kaplan and Meystre(1981)Kaplan, and Meystre]{kaplan1981enhancement}
Kaplan,~A.; Meystre,~P. Enhancement of the Sagnac effect due to nonlinearly
  induced nonreciprocity. \emph{Optics Letters} \textbf{1981}, \emph{6},
  590--592\relax
\mciteBstWouldAddEndPuncttrue
\mciteSetBstMidEndSepPunct{\mcitedefaultmidpunct}
{\mcitedefaultendpunct}{\mcitedefaultseppunct}\relax
\EndOfBibitem
\bibitem[Kaplan and Meystre(1982)Kaplan, and Meystre]{kaplan1982directionally}
Kaplan,~A.; Meystre,~P. Directionally asymmetrical bistability in a
  symmetrically pumped nonlinear ring interferometer. \emph{Optics
  Communications} \textbf{1982}, \emph{40}, 229--232\relax
\mciteBstWouldAddEndPuncttrue
\mciteSetBstMidEndSepPunct{\mcitedefaultmidpunct}
{\mcitedefaultendpunct}{\mcitedefaultseppunct}\relax
\EndOfBibitem
\bibitem[Kaplan(1983)]{kaplan1983light}
Kaplan,~A. Light-induced nonreciprocity, field invariants, and nonlinear
  eigenpolarizations. \emph{Optics Letters} \textbf{1983}, \emph{8},
  560--562\relax
\mciteBstWouldAddEndPuncttrue
\mciteSetBstMidEndSepPunct{\mcitedefaultmidpunct}
{\mcitedefaultendpunct}{\mcitedefaultseppunct}\relax
\EndOfBibitem
\bibitem[Pozar(2011)]{pozar2021microwave}
Pozar,~D.~M. \emph{Microwave engineering: theory and techniques}, 4th ed.; John
  wiley \& sons, 2011\relax
\mciteBstWouldAddEndPuncttrue
\mciteSetBstMidEndSepPunct{\mcitedefaultmidpunct}
{\mcitedefaultendpunct}{\mcitedefaultseppunct}\relax
\EndOfBibitem
\bibitem[Barron(1986)]{barron1986true}
Barron,~L.~D. True and false chirality and absolute asymmetric synthesis.
  \emph{Journal of the American Chemical Society} \textbf{1986}, \emph{108},
  5539--5542\relax
\mciteBstWouldAddEndPuncttrue
\mciteSetBstMidEndSepPunct{\mcitedefaultmidpunct}
{\mcitedefaultendpunct}{\mcitedefaultseppunct}\relax
\EndOfBibitem
\bibitem[Barron(1986)]{barron1986true2}
Barron,~L.~D. True and false chirality and parity violation. \emph{Chemical
  Physics Letters} \textbf{1986}, \emph{123}, 423--427\relax
\mciteBstWouldAddEndPuncttrue
\mciteSetBstMidEndSepPunct{\mcitedefaultmidpunct}
{\mcitedefaultendpunct}{\mcitedefaultseppunct}\relax
\EndOfBibitem
\bibitem[Barron(1986)]{barron1986symmetry}
Barron,~L. Symmetry and molecular chirality. \emph{Chemical Society Reviews}
  \textbf{1986}, \emph{15}, 189--223\relax
\mciteBstWouldAddEndPuncttrue
\mciteSetBstMidEndSepPunct{\mcitedefaultmidpunct}
{\mcitedefaultendpunct}{\mcitedefaultseppunct}\relax
\EndOfBibitem
\bibitem[de~Figueiredo and Raab(1980)de~Figueiredo, and Raab]{de1980pictorial}
de~Figueiredo,~I.~M.; Raab,~R. A pictorial approach to macroscopic space-time
  symmetry, with particular reference to light scattering. \emph{Proceedings of
  the Royal Society of London. A. Mathematical and Physical Sciences}
  \textbf{1980}, \emph{369}, 501--516\relax
\mciteBstWouldAddEndPuncttrue
\mciteSetBstMidEndSepPunct{\mcitedefaultmidpunct}
{\mcitedefaultendpunct}{\mcitedefaultseppunct}\relax
\EndOfBibitem
\bibitem[Bliokh \latin{et~al.}(2014)Bliokh, Kivshar, and
  Nori]{bliokh2014magnetoelectric}
Bliokh,~K.~Y.; Kivshar,~Y.~S.; Nori,~F. Magnetoelectric effects in local
  light-matter interactions. \emph{Physical review letters} \textbf{2014},
  \emph{113}, 033601\relax
\mciteBstWouldAddEndPuncttrue
\mciteSetBstMidEndSepPunct{\mcitedefaultmidpunct}
{\mcitedefaultendpunct}{\mcitedefaultseppunct}\relax
\EndOfBibitem
\bibitem[Hassani~Gangaraj and Monticone(2022)Hassani~Gangaraj, and
  Monticone]{hassani2022drifting}
Hassani~Gangaraj,~S.~A.; Monticone,~F. Drifting electrons: nonreciprocal
  plasmonics and thermal photonics. \emph{ACS photonics} \textbf{2022},
  \emph{9}, 806--819\relax
\mciteBstWouldAddEndPuncttrue
\mciteSetBstMidEndSepPunct{\mcitedefaultmidpunct}
{\mcitedefaultendpunct}{\mcitedefaultseppunct}\relax
\EndOfBibitem
\bibitem[Hassani~Gangaraj \latin{et~al.}(2019)Hassani~Gangaraj, Hanson,
  Silveirinha, Shastri, Antezza, and Monticone]{hassani2019unidirectional}
Hassani~Gangaraj,~S.~A.; Hanson,~G.~W.; Silveirinha,~M.~G.; Shastri,~K.;
  Antezza,~M.; Monticone,~F. Unidirectional and diffractionless surface plasmon
  polaritons on three-dimensional nonreciprocal plasmonic platforms.
  \emph{Physical Review B} \textbf{2019}, \emph{99}, 245414\relax
\mciteBstWouldAddEndPuncttrue
\mciteSetBstMidEndSepPunct{\mcitedefaultmidpunct}
{\mcitedefaultendpunct}{\mcitedefaultseppunct}\relax
\EndOfBibitem
\bibitem[Guo and Argyropoulos(2022)Guo, and Argyropoulos]{guo2022nonreciprocal}
Guo,~T.; Argyropoulos,~C. Nonreciprocal transmission of electromagnetic waves
  with nonlinear active plasmonic metasurfaces. \emph{Physical Review B}
  \textbf{2022}, \emph{106}, 235418\relax
\mciteBstWouldAddEndPuncttrue
\mciteSetBstMidEndSepPunct{\mcitedefaultmidpunct}
{\mcitedefaultendpunct}{\mcitedefaultseppunct}\relax
\EndOfBibitem
\bibitem[Pancharatnam(1956)]{pancharatnam1956generalized1}
Pancharatnam,~S. Generalized theory of interference, and its applications: Part
  I. Coherent pencils. Proceedings of the Indian Academy of Sciences-Section A.
  1956; pp 247--262\relax
\mciteBstWouldAddEndPuncttrue
\mciteSetBstMidEndSepPunct{\mcitedefaultmidpunct}
{\mcitedefaultendpunct}{\mcitedefaultseppunct}\relax
\EndOfBibitem
\bibitem[Pancharatnam(1956)]{pancharatnam1956generalized}
Pancharatnam,~S. Generalized theory of interference and its applications: Part
  II. Partially coherent pencils. Proceedings of the Indian Academy of
  Sciences-section a. 1956; pp 398--417\relax
\mciteBstWouldAddEndPuncttrue
\mciteSetBstMidEndSepPunct{\mcitedefaultmidpunct}
{\mcitedefaultendpunct}{\mcitedefaultseppunct}\relax
\EndOfBibitem
\bibitem[Berry(1987)]{berry1987adiabatic}
Berry,~M.~V. The adiabatic phase and Pancharatnam's phase for polarized light.
  \emph{Journal of Modern Optics} \textbf{1987}, \emph{34}, 1401--1407\relax
\mciteBstWouldAddEndPuncttrue
\mciteSetBstMidEndSepPunct{\mcitedefaultmidpunct}
{\mcitedefaultendpunct}{\mcitedefaultseppunct}\relax
\EndOfBibitem
\bibitem[Jisha \latin{et~al.}(2021)Jisha, Nolte, and
  Alberucci]{jisha2021geometric}
Jisha,~C.~P.; Nolte,~S.; Alberucci,~A. Geometric phase in optics: from
  wavefront manipulation to waveguiding. \emph{Laser \& Photonics Reviews}
  \textbf{2021}, \emph{15}, 2100003\relax
\mciteBstWouldAddEndPuncttrue
\mciteSetBstMidEndSepPunct{\mcitedefaultmidpunct}
{\mcitedefaultendpunct}{\mcitedefaultseppunct}\relax
\EndOfBibitem
\bibitem[Cohen \latin{et~al.}(2019)Cohen, Larocque, Bouchard, Nejadsattari,
  Gefen, and Karimi]{cohen2019geometric}
Cohen,~E.; Larocque,~H.; Bouchard,~F.; Nejadsattari,~F.; Gefen,~Y.; Karimi,~E.
  Geometric phase from Aharonov--Bohm to Pancharatnam--Berry and beyond.
  \emph{Nature Reviews Physics} \textbf{2019}, \emph{1}, 437--449\relax
\mciteBstWouldAddEndPuncttrue
\mciteSetBstMidEndSepPunct{\mcitedefaultmidpunct}
{\mcitedefaultendpunct}{\mcitedefaultseppunct}\relax
\EndOfBibitem
\bibitem[Samuel and Bhandari(1988)Samuel, and Bhandari]{samuel1988general}
Samuel,~J.; Bhandari,~R. General setting for Berry's phase. \emph{Physical
  Review Letters} \textbf{1988}, \emph{60}, 2339\relax
\mciteBstWouldAddEndPuncttrue
\mciteSetBstMidEndSepPunct{\mcitedefaultmidpunct}
{\mcitedefaultendpunct}{\mcitedefaultseppunct}\relax
\EndOfBibitem
\bibitem[De~Zela(2012)]{de2012pancharatnam}
De~Zela,~F. \emph{The pancharatnam-berry phase: Theoretical and experimental
  aspects}; InTech Rijeka, Croatia, 2012\relax
\mciteBstWouldAddEndPuncttrue
\mciteSetBstMidEndSepPunct{\mcitedefaultmidpunct}
{\mcitedefaultendpunct}{\mcitedefaultseppunct}\relax
\EndOfBibitem
\bibitem[Lor{\'e}n \latin{et~al.}(2023)Lor{\'e}n, Paravicini-Bagliani, Saha,
  Gautier, Li, Genet, and Mart{\'\i}n-Moreno]{loren2023microscopic}
Lor{\'e}n,~F.; Paravicini-Bagliani,~G.~L.; Saha,~S.; Gautier,~J.; Li,~M.;
  Genet,~C.; Mart{\'\i}n-Moreno,~L. Microscopic analysis of spin-momentum
  locking on a geometric phase metasurface. \emph{Physical Review B}
  \textbf{2023}, \emph{107}, 165128\relax
\mciteBstWouldAddEndPuncttrue
\mciteSetBstMidEndSepPunct{\mcitedefaultmidpunct}
{\mcitedefaultendpunct}{\mcitedefaultseppunct}\relax
\EndOfBibitem
\bibitem[Fox and Gorodetski(2022)Fox, and Gorodetski]{fox2022generalized}
Fox,~M.; Gorodetski,~Y. Generalized approach to plasmonic phase modulation in
  topological bi-gratings. \emph{Applied Physics Letters} \textbf{2022},
  \emph{120}, 031105\relax
\mciteBstWouldAddEndPuncttrue
\mciteSetBstMidEndSepPunct{\mcitedefaultmidpunct}
{\mcitedefaultendpunct}{\mcitedefaultseppunct}\relax
\EndOfBibitem
\bibitem[Culshaw(2005)]{culshaw2005optical}
Culshaw,~B. The optical fibre Sagnac interferometer: an overview of its
  principles and applications. \emph{Measurement Science and Technology}
  \textbf{2005}, \emph{17}, R1\relax
\mciteBstWouldAddEndPuncttrue
\mciteSetBstMidEndSepPunct{\mcitedefaultmidpunct}
{\mcitedefaultendpunct}{\mcitedefaultseppunct}\relax
\EndOfBibitem
\bibitem[Homola(2003)]{homola2003present}
Homola,~J. Present and future of surface plasmon resonance biosensors.
  \emph{Analytical and Bioanalytical Chemistry} \textbf{2003}, \emph{377},
  528--539\relax
\mciteBstWouldAddEndPuncttrue
\mciteSetBstMidEndSepPunct{\mcitedefaultmidpunct}
{\mcitedefaultendpunct}{\mcitedefaultseppunct}\relax
\EndOfBibitem
\bibitem[Fox and Gorodetski(2023)Fox, and Gorodetski]{fox2023topologically}
Fox,~M.; Gorodetski,~Y. Topologically protected plasmonic phases in randomized
  aperture gratings. \emph{Scientific Reports} \textbf{2023}, \emph{13},
  1006\relax
\mciteBstWouldAddEndPuncttrue
\mciteSetBstMidEndSepPunct{\mcitedefaultmidpunct}
{\mcitedefaultendpunct}{\mcitedefaultseppunct}\relax
\EndOfBibitem
\bibitem[Lor{\'e}n \latin{et~al.}(2023)Lor{\'e}n, Genet, and
  Mart{\'\i}n-Moreno]{loren2023spinmomentum}
Lor{\'e}n,~F.; Genet,~C.; Mart{\'\i}n-Moreno,~L. Spin-momentum locking
  breakdown on plasmonic metasurfaces. \emph{Physical Review B} \textbf{2023},
  \emph{108}, 155144\relax
\mciteBstWouldAddEndPuncttrue
\mciteSetBstMidEndSepPunct{\mcitedefaultmidpunct}
{\mcitedefaultendpunct}{\mcitedefaultseppunct}\relax
\EndOfBibitem
\bibitem[Bomzon \latin{et~al.}(2001)Bomzon, Kleiner, and
  Hasman]{bomzon2001pancharatnam}
Bomzon,~Z.; Kleiner,~V.; Hasman,~E. Pancharatnam--Berry phase in space-variant
  polarization-state manipulations with subwavelength gratings. \emph{Optics
  Letters} \textbf{2001}, \emph{26}, 1424--1426\relax
\mciteBstWouldAddEndPuncttrue
\mciteSetBstMidEndSepPunct{\mcitedefaultmidpunct}
{\mcitedefaultendpunct}{\mcitedefaultseppunct}\relax
\EndOfBibitem
\bibitem[Gorodetski \latin{et~al.}(2012)Gorodetski, Bliokh, Stein, Genet,
  Shitrit, Kleiner, Hasman, and Ebbesen]{gorodetski2012weak}
Gorodetski,~Y.; Bliokh,~K.; Stein,~B.; Genet,~C.; Shitrit,~N.; Kleiner,~V.;
  Hasman,~E.; Ebbesen,~T. Weak measurements of light chirality with a plasmonic
  slit. \emph{Physical Review Letters} \textbf{2012}, \emph{109}, 013901\relax
\mciteBstWouldAddEndPuncttrue
\mciteSetBstMidEndSepPunct{\mcitedefaultmidpunct}
{\mcitedefaultendpunct}{\mcitedefaultseppunct}\relax
\EndOfBibitem
\bibitem[Collett(1992)]{collett1992polarized}
Collett,~E. \emph{Polarized light. Fundamentals and applications}; New York :
  Marcel Dekker, 1992\relax
\mciteBstWouldAddEndPuncttrue
\mciteSetBstMidEndSepPunct{\mcitedefaultmidpunct}
{\mcitedefaultendpunct}{\mcitedefaultseppunct}\relax
\EndOfBibitem
\end{mcitethebibliography}


\providecommand{\latin}[1]{#1}
\makeatletter
\providecommand{\doi}
  {\begingroup\let\do\@makeother\dospecials
  \catcode`\{=1 \catcode`\}=2 \doi@aux}
\providecommand{\doi@aux}[1]{\endgroup\texttt{#1}}
\makeatother
\providecommand*\mcitethebibliography{\thebibliography}
\csname @ifundefined\endcsname{endmcitethebibliography}
  {\let\endmcitethebibliography\endthebibliography}{}
\begin{mcitethebibliography}{0}
\providecommand*\natexlab[1]{#1}
\providecommand*\mciteSetBstSublistMode[1]{}
\providecommand*\mciteSetBstMaxWidthForm[2]{}
\providecommand*\mciteBstWouldAddEndPuncttrue
  {\def\EndOfBibitem{\unskip.}}
\providecommand*\mciteBstWouldAddEndPunctfalse
  {\let\EndOfBibitem\relax}
\providecommand*\mciteSetBstMidEndSepPunct[3]{}
\providecommand*\mciteSetBstSublistLabelBeginEnd[3]{}
\providecommand*\EndOfBibitem{}
\mciteSetBstSublistMode{f}
\mciteSetBstMaxWidthForm{subitem}{(\alph{mcitesubitemcount})}
\mciteSetBstSublistLabelBeginEnd
  {\mcitemaxwidthsubitemform\space}
  {\relax}
  {\relax}

\end{mcitethebibliography}
\providecommand{\latin}[1]{#1}
\makeatletter
\providecommand{\doi}
{\begingroup\let\do\@makeother\dospecials
	\catcode`\{=1 \catcode`\}=2 \doi@aux}
\providecommand{\doi@aux}[1]{\endgroup\texttt{#1}}
\makeatother
\providecommand*\mcitethebibliography{\thebibliography}
\csname @ifundefined\endcsname{endmcitethebibliography}
{\let\endmcitethebibliography\endthebibliography}{}

\end{document}